\documentclass[fleqn,12pt]{wlscirep}
\usepackage[utf8]{inputenc}
\usepackage[T1]{fontenc}
\usepackage{bm}

\usepackage{setspace}
\usepackage{fontsize}
\usepackage{xcolor}

\usepackage{mathtools}
\usepackage{amssymb}
\usepackage{braket}

\usepackage[justification=justified]{caption}
\usepackage[none]{hyphenat}

\newcommand{\photonweight}{\left|\alpha_{ph}\right|^2}

\title{
Unveiling the mixed nature of polaritonic transport: From enhanced diffusion to ballistic motion approaching the speed of light
}

\author[1]{M Balasubrahmaniyam}
\author[1]{Arie Simkovich}
\author[1]{Adina Golombek}
\author[2]{Guy Ankonina}
\author[1,*]{Tal Schwartz}
\affil[1]{School of Chemistry, Raymond and Beverly Sackler Faculty of Exact Sciences and Tel Aviv University Center for Light-Matter Interaction, Tel Aviv University, Tel Aviv 6997801, Israel.}
\affil[2]{Russell Berrie Nanotechnology Institute, Technion, Haifa 3200003, Israel}

\affil[*]{e-mail: talschwartz@tau.ac.il}

\begin{abstract}
\changefontsize[11pt]{11pt}
In recent years it has become clear that the transport of excitons and charge carriers in molecular systems can be enhanced by coherent coupling with photons, giving rise to the formation of hybrid excitations known as polaritons.
Such enhancement has far-reaching technological implications, however, the enhancement mechanism and the transport nature of these composite light-matter excitations in such systems still remain elusive.
Here we map the ultrafast spatiotemporal dynamics of surface-bound optical waves strongly coupled to a self-assembled molecular layer and fully resolve them in energy/momentum space.
Our studies reveal intricate behavior which stems from the hybrid nature of polaritons.
We find that the balance between the molecular disorder and long-range correlations induced by the coherent mixing between light and matter leads to a mobility transition between diffusive and ballistic transport, which can be controlled by varying the light-matter composition of the polaritons.
Furthermore, we directly demonstrate that the coupling with light can enhance the diffusion coefficient of molecular excitons by six orders of magnitude and even lead to ballistic flow at two-thirds the speed of light.
\end{abstract}

\begin{document}

\setstretch{2}
\changefontsize[16pt]{12pt}

\flushbottom
\maketitle

\thispagestyle{empty}

\section*{Introduction}

Transport of energy and charge carriers in organic materials and molecular systems is fundamental to various fields ranging from photochemistry to optoelectronics.
It is essential to photosynthesis~\cite{Scholes2011} and governs the operation of organic solar cells and other organic electronic devices~\cite{Forrest2020, Tessler2009}.
However, due to the highly disordered nature of such materials, transport in organic materials is dominated by short-range random hopping, with diffusion length values typically in the nanometric, molecular scale and reaching hundreds of nanometers at most~\cite{Ginsberg2020}.
This leads to slow and inefficient transport, which is typically diffusive or sub-diffusive, and often poses severe technological limitations when using organic materials in applications.
Recently, it has been demonstrated that both excitonic and electronic transport in organic semiconductors can be enhanced by coupling these materials with the photonic modes of an optical resonator \cite{Orgiu2015,Lerario2017,Rozenman2018,Zakharko2018,Nagarajan2020, Hou2020,Bhatt2021,Pandya2021,Pandya2022}.
Under suitable conditions, collective strong coupling between the material excitations and light can give rise to hybrid excitations called polaritons, which are partly excitonic and partly photonic~\cite{Francisco2021}.
As such, the polaritonic wavefunctions can extend over length-scales exceeding the optical wavelength and cover a macroscopically large number of molecules.
This, in turn, results in quantum correlations between remote molecules\cite{AberraGuebrou2012}, which can facilitate energy transfer~\cite{Zhong2017,Du2018,Georgiou2021} and transport phenomena.
This kind of cavity-enhanced transport has been drawing increasing interest over the past few years, with both experimental \cite{Freixanet2000,Orgiu2015,Lerario2017,Rozenman2018,Zakharko2018,Paravicini-Bagliani2018,Thomas2019,Nagarajan2020,Hou2020,Bhatt2021,Pandya2021,Pandya2022} and theoretical\cite{Feist2015,Schachenmayer2015,Gonzalez-Ballestero2016,Hagenmuller2017,Sentef2018,Botzung2020,Chavez2021,Engelhardt2022} efforts devoted to understanding how strong coupling affects transport phenomena.
Although the long-range propagation of polaritons has been directly visualized and studied in steady-state experiments \cite{Lerario2017,Zakharko2018,Hou2020}, constructing a complete picture of their transport dynamics necessitates access to the kinetics of the polaritonic motion, in a similar manner to the various time-resolved microscopy techniques used for studying normal organic semiconductors \cite{Virgili2012,Akselrod2014,Zhu2017,Wan2015,Delor2020,Berghuis2021}.
Indeed, by employing ultrafast microscopy, a few recent experiments successfully resolved the spatiotemporal evolution of polaritons \cite{Rozenman2018,Pandya2022}. In particular, we have observed how polaritons, excited in a metallic, planar cavity, gradually migrate and reach distances which are $\sim 5~\mu$m away from their initial position, within several picoseconds.
Interestingly, these studies found that the polariton velocity was much lower than expected by simple considerations\cite{Rozenman2018,Pandya2022}.
While the dispersion of cavity polaritons predicts a group velocity on the order of 20~$\mu$m$/$ps, the observed velocities were smaller than 1~$\mu$m$/$ps, pointing toward a fundamental gap in the understanding of cavity polaritons.
Furthermore, even today the nature of polaritonic transport is still obscure - although the coupling of the excitons to propagating photons is assumed to lead to ballistic motion, these few initial studies suggest that this is not the case in reality, and that disorder, which is inherent to all molecular systems, renders the transport of polaritons non-ballistic.

To close this gap and to enable the practical implementation of cavity-enhanced transport in future applications, we study the ultrafast spatiotemporal dynamics in a polaritonic system based on a fully-dielectric, 1D photonic crystal which supports surface-bound polaritons with ultra-long propagation lengths of $\sim 100~\mu$m.
Our measurements fully reveal the intricate details of polaritonic transport in molecular systems and how it depends on the balance between the photonic and excitonic components comprising the polaritons. 
As we show here for the first time, polaritons exhibit a mobility transition between diffusive and ballistic transport, which occurs as these collective molecular excitations become more photon-like.
Furthermore, our measurements allow us to obtain, in the most direct manner, the various parameters that govern the motion of polaritons.
We find that the strong coupling with light and the long-range correlations counteract the molecular disorder, boosting the diffusion coefficient by a factor of $10^6$, and even completely overcome disorder, leading to ballistic flow.
This ballistic propagation is shown to proceed over macroscopic distances of $\sim 100~\mu$m and at two thirds the speed of light in vacuum.

\section*{Bloch Surface Wave Polaritons: Steady State Characterization}

\begin{figure}[htp]
\centering
\includegraphics[width=175mm]{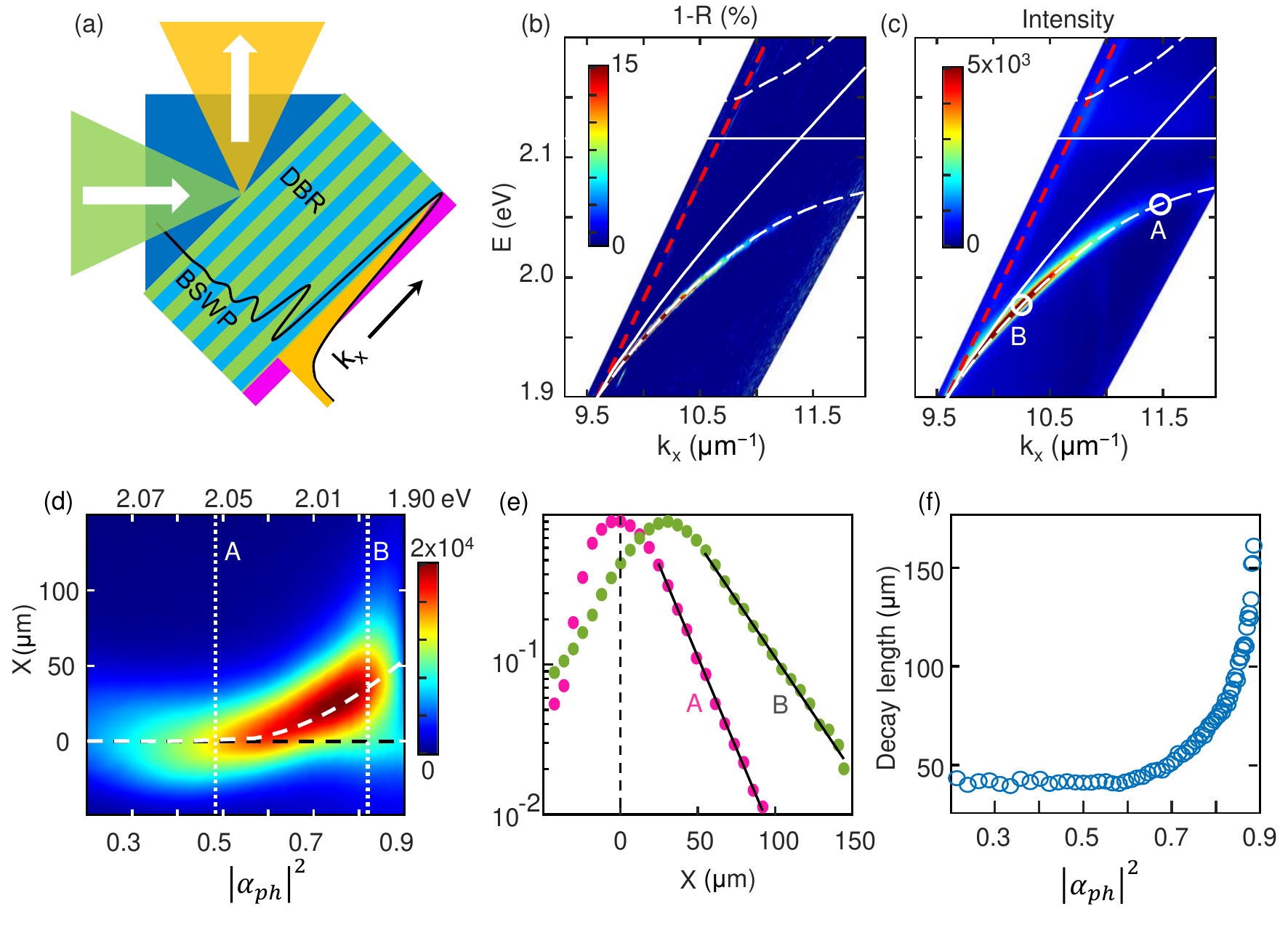}
\caption{
    (a) Illustration of the DBR structure (in alternating green and blue) coated with TDBC molecules and probed in Kretschmann (prism-coupled) configuration.
    The black curve shows the simulated field distribution of the excited BSWP modes.
    (b,c) Dispersion of the BSWP system measured via angle-resolved reflection (b) and emission (c) spectroscopy (represented by the false-color plots).
    The sharp signal under the lower white dashed lines corresponds to the lower BSW polariton mode, while the white dashed lines represent the simulated dispersion (see Supplementary Information).
    The solid white lines indicate the bare BSW dispersion and the exciton energy (fixed at 2.13 eV), while the red dashed line indicates the light line.
    (d) Steady state spatial emission distribution of the lower BSWP as a function of its photonic weight, measured with a localized nonresonant excitation of the polaritons at $x=0$ (indicated by the black dashed line). The white dashed curve marks the variation of the emission peak position with the photonic weight, showing the migration of the BSWP's in space.
    (e) Representative steady state emission profiles (note logarithmic scale), measured at the points A and B indicated in (c) and (d).
    The black solid lines show the exponential fits to the data.
    (f) Emission decay length as a function of photonic weight, extracted from the exponential fit to the tails of the steady-state distributions given in (d).
}
\label{fig:ss}
\end{figure}

The system studied in this work consists of a distributed Bragg reflector (DBR) structure, made of alternating layers of TiO$_2$ and SiO$_2$, and covered with three monolayers of organic semiconductor TDBC molecules (see Supplementary Information for details of the system and preparation methods).
As depicted in Fig.~\ref{fig:ss}(a), the DBR structure supports a TE-polarized surface wave mode (Bloch surface wave, BSW) within its bandgap \cite{Yeh1978}, which can interact with the molecular film to form delocalized BSW polaritons (BSWP) in the strong coupling regime \cite{Lerario2014,Lerario2017,Hou2020}.
To characterize the steady-state (linear) properties of the BSWP we utilize a dedicated spectral imaging setup designed for either angle resolved or spatially resolved reflection/emission measurements, implemented in a Kretschmann configuration as shown in Fig.~\ref{fig:ss}(a) (for details see Supplementary Information and Ref. \citen{Ohad2018}).
From the angle-resolved spectroscopy we obtain the dispersion curve of the resulting polaritons, which is seen in Fig.~\ref{fig:ss}(b) (reflection) and \ref{fig:ss}(c) (emission), residing beyond the air light line (marked by red lines).
The bare TDBC absorption line at 2.13 eV (582 nm) is marked by the horizontal white line, while the dashed white lines correspond to the simulated dispersion of the lower and upper BSWP modes, exhibiting excellent agreement with the experimental results.
From the measured dispersion we deduce a Rabi splitting energy of 142 meV at the resonant point, which occurs at an angle of incidence of $\theta=45^\circ$ (when probing the sample from the prism side.
Furthermore, by examining the linewidth of the polariton dispersion (see Supplementary Information) we obtain a Q-factor of 160 at 2.06 eV (at the anti-crossing point), which increases to $\sim$ 400 for BSWP modes residing far below the molecular absorption.

Before proceeding to the dynamical measurements, we characterize the polariton propagation by examining the steady-state spatial distribution of polaritons following a localized, non-resonant excitation (see Supplementary Information for details), in a similar manner to previous studies \cite{Lerario2017,Hou2020}.
We record the steady state emission profile of the BSWP for various wavelengths and plot it in fig.\ref{fig:ss}(d).
Here, instead of the polariton energy, it is more meaningful to examine the polariton propagation as a function of its photon-like character, which is quantified by the (energy-dependent) photonic weight, given by $\photonweight = (E_x-E_p)^2/\left[g^2+(E_x-E_p)^2\right]$, where $g=71$~meV is the coupling constant (half of the Rabi-splitting energy),~$E_x= 2.13$ eV is the molecular exciton energy for TDBC and $E_p (k_x)$ is the polariton energy.
Note that the photonic weight of the polariton increases as it shifts from the exciton energy, such that for the lower BSWP branch the polaritons become more photon-like as their energy decreases (see Supplementary Information).

As seen, the spatial distribution of the polaritonic emission exhibits both broadening as well as shifting of the peak position, both of which become more pronounced as the photonic weight increases, reaching a scale of $50-100~\mu $m.
It is important to note that only emission from BSWPs propagating in the positive $x$ direction will be routed toward the imaging system by the prism-coupler, which is essentially analogous to filtering in the Fourier domain.
Such filtering leads to the apparent asymmetry in the emission profiles seen in Fig.~\ref{fig:ss}(d)\cite{Lerario2017,Hou2020}.
In Fig.~\ref{fig:ss}(e) we plot (in logarithmic scale) two representative intensity profiles (i.e. vertical cross sections of Fig.~\ref{fig:ss}(d) taken at points A and B), with photonic weights of $ \photonweight =0.48\ $ and $\photonweight =0.82$.
Both of these spatial distributions clearly show an exponentially decaying tale.
However, the decay length at point B, for which the polariton has a larger photonic component, is larger than that observed at point A.
Indeed, when repeating the exponential fit to the entire data shown in Fig.~\ref{fig:ss}(d) and plotting the decay length as a function of the photonic weight, we clearly see a monotonic increase  of decay length, reaching distances of more than 150 $\mu$m. This confirms that ultra-long propagation takes place in the system, as expected from previous steady-state studies of BSWP\cite{Lerario2017,Hou2020}.
Nevertheless, as in those previous studies, such steady-state measurements do not provide any information regarding the propagation velocities, the transport mechanism or any other dynamical property of the polariton motion.

\section*{Dynamics Measurements and Results}
\begin{figure}[ht]
    \centering
    \includegraphics[width=175mm]{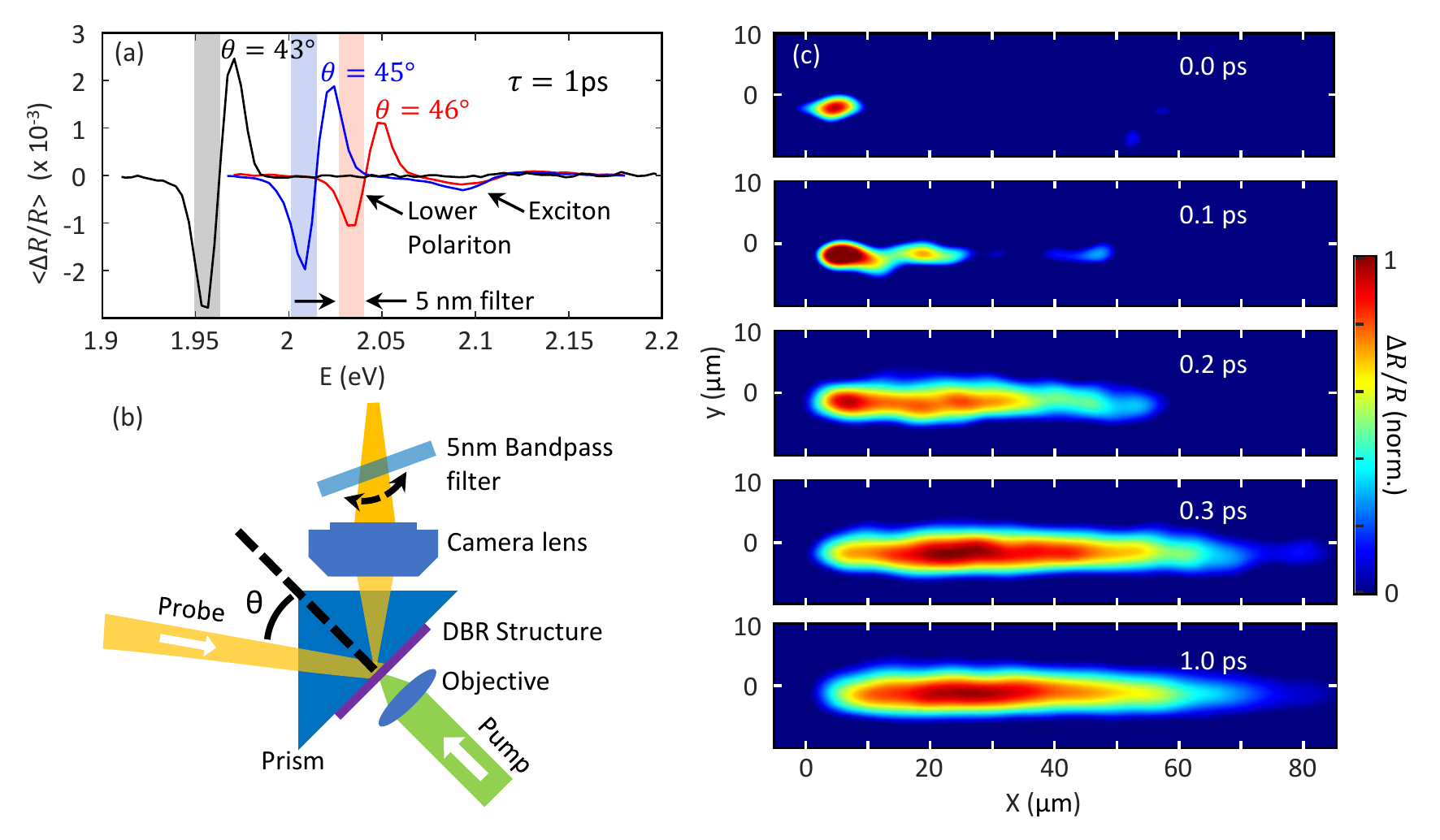}
    \caption{
    (a) Transient reflection spectra captured at $\tau=1$ ps for probe incident angles of $\theta \sim 43^{\circ}$, $45^{\circ}$ and $46^{\circ}$.
    The shaded regions mark the 5 nm-wide spectral band which is probed by the ultrafast microscopy setup (see variable band-pass filter in (b)).
    (b) A schematic sketch of the optical configuration used in the pump-probe microscopy experiments (see Supplementary Information for full details).
    (c) Representative snapshots obtained from the time-resolved microscopy for $\photonweight =0.82$ showing the gradual expansion of the polariton cloud.
    The data correspond to the spatial distribution of the (normalized) differential reflectivity, probed at an angle of incidence of $\theta = 43^{\circ}$ and with the 5~nm bandpass filter centered around $1.955$~eV as indicated by the gray shaded region in the pump-probe spectra in (a).
    }
    \label{fig:pump-probe}
\end{figure}

To reveal the full transport dynamics of the polaritons, we proceed to studying their kinetics, in particular, their spatiotemporal dynamics and how these depend on the polariton composition.
We start by measuring the differential reflection spectrum (averaged over space, see Supplementary Information) for three different angles of incidence of the probe ($\theta \simeq 43^\circ$, $45^\circ$ and $46^\circ$).
The spectra, plotted in Fig.~\ref{fig:pump-probe}(a) for a time-delay of $1$ psec, show two resonant features, similar to the transient spectra observed in strongly coupled Fabry-P\'erot cavities\cite{Schwartz2013}: a prominent, angle-dependent feature around the energy of the lower BSWP (corresponding to $ \photonweight $ values of 0.82, 0.66 and 0.54) and a second, weaker one that occurs at the bare exciton energy and does not show any angular dependence. The measured decay rates for the BSWP's are 3.8, 6.5 and 6.6 ps respectively while at the bare exciton energy we observe a decay rate of 6.6 (see Supplementary Information).

To capture the full spatiotemporal dynamics of BSWPs we use a custom-built  ultrafast pump-probe microscopy system \cite{Rozenman2018,Fischer2016}, a sketch of which is presented in Fig.~\ref{fig:pump-probe}(b) (see Supplementary Information for details of the system and measurements).
We operate the pump-probe setup in the Kretschmann attenuated total internal reflection configuration, which allows for measurement beyond the air light line, where the BSWPs reside.
The pump and probe beams are spatially overlapping in the plane of the sample, with the probed area diameter ($\sim 1$mm) being much larger than both the imaging field of view ($\sim 90\mu$m) and the the area excited by the pump beam (having a diameter of $\sim 5\mu$m), in order to measure the expanding polariton cloud in a wide-field (i.e. nonscanning) mode.

The nonresonant excitation of the system is followed by a fast energy relaxation process which distributes the energy among all the various polaritonic modes, filling the entire lower BSWP branch within a few tens of femtoseconds\cite{Schwartz2013,Mewes2020}.
However, in order to resolve the transport dynamics of a particular polaritonic excitation (i.e. at a specific location along the lower BSWP dispersion curve in Fig.~\ref{fig:ss}(b)), we need to match the probe energy and in-plane momentum to those of the polariton.
The probe energy is tuned by a band-pass filter, whose transmission band is set to overlap with the negative spectral feature of the corresponding polariton (represented by the shaded gray area in Fig.~\ref{fig:pump-probe}(a)). At the same time, the probe in-plane momentum is selected by varying its angle of incidence as shown in \ref{fig:pump-probe}(b) (see Supplementary Information for details).

In Fig.~\ref{fig:pump-probe}(c) we plot a representative sequence of pump-probe images (i.e. differential reflectivity $\Delta R/R$) measured at $E=1.96$ eV (corresponding to BSWP with $ \photonweight = 0.82 $) and at different delay times.
The signal measured here is directly proportional to the local density of excited molecules, and hence these images capture the spatial distribution of the expanding polariton cloud, providing direct evidence of the enhanced energy transport over macroscopic length-scales.
Starting out with a Gaussian-like profile with a full width at half maximum (FWHM) of $5\mu$m at 0 ps delay, the polariton cloud expands along the $x$ direction (matching the direction of the in-plane component of the probe) to a width of $58 \mu$m  within 0.3 ps.
These results show a similar trend to our previous time-resolved imaging experiments using a Fabry-P\'erot cavity\cite{Rozenman2018}, however, here we observe propagation distances which are an order of magnitude longer (as also seen in our steady-state measurements, Fig.~\ref{fig:ss}(d)).
Moreover, a rough estimate of the expansion velocity gives a value of $\sim 176~\mu$m$/$ps, roughly two thirds the speed of light.
This value is more than two orders of magnitude larger than observed for polaritons in any previous time-resolved measurements conducted in Fabry-P\'erot cavity systems \cite{Freixanet2000,Rozenman2018, Pandya2022}. In fact, this estimated velocity approaches the polariton group velocity at the same location on the lower BSWP branch (point B in Fig.~\ref{fig:ss}(c)), which is evaluated as $\sim 190 \mu m/ps $.

\begin{figure}[ht]
\centering
\includegraphics[width=175mm]{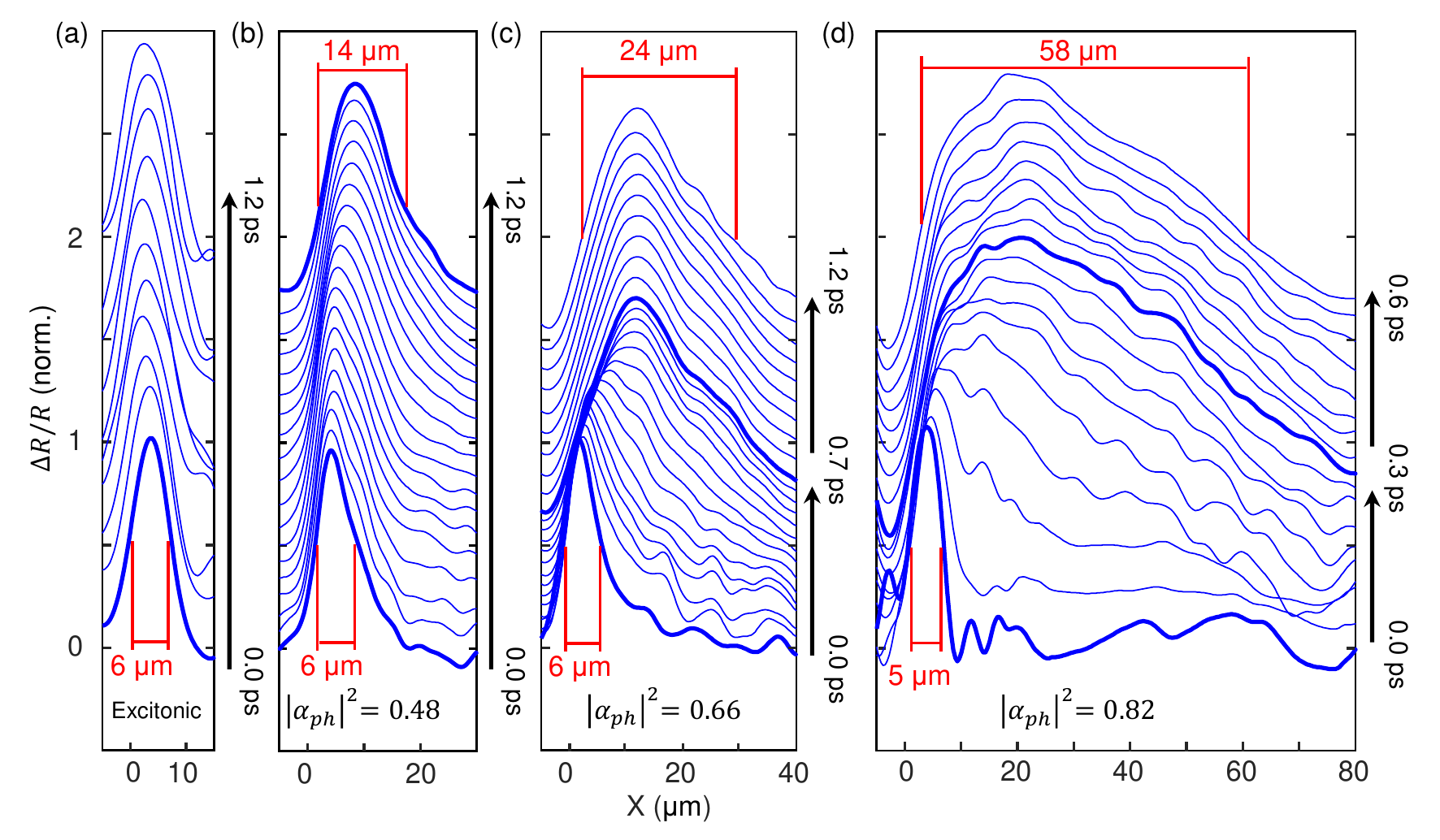}
\caption{
    (a) Horizontal cross sections of the pump-probe signal measured at the bare exciton energy (2.13~eV) as a function of the pump-probe delay (with the increase in delay time indicated by the black arrows).
    (b)-(d) Cross-sections of the BSWP spatial distributions as a function of the pump-probe delay, demonstrating the gradual expansion of the polariton cloud.
    The dynamics in (b), (c) and (d) were measured at three different points along the BSWP dispersion with photonic weights $\photonweight =$0.48, 0.66 and 0.82 respectively.
    The cross-sections are normalised and shifted vertically for visibility.}
\label{fig:dynamics}
\end{figure}

The detailed dynamics of the BSWP cloud as measured by the pump-probe microscopy are provided in Fig.~\ref{fig:dynamics}, which shows the cross-sections of the (normalised) differential signals.
Fig.~\ref{fig:dynamics}(a) shows the cross-sections of the signal measured at the bare-exciton energy (2.13 eV or 582 nm), whereas (b), (c) and (d) show the evolution of the BSWP cloud for $\photonweight$ = 0.48, 0.66 and 0.82 respectively.
At the bare exciton energy, corresponding to uncoupled excitons/dark states, the width (quantified as twice the standard deviation of the intensity profile, see see Supplementary Information) remains approximately 6 $\mu m$ for all delays.
This result is expected, since the typical diffusion length of these excitations is much shorter than spatial resolution in our measurement.
In sharp contrast, when we tune the probe to match the BSWP state with $\photonweight = 0.48 $ (i.e. similar photonic and excitonic weights) the polariton cloud expands with time, until reaching a steady state width of $ \sim 14 \mu$m at $\sim 1.2 $ ps, after which  the profile does not expand but merely exhibits a spatially uniform decay with a lifetime of $\sim 6.8$~ps (see Supplementary Information).
As the photonic weight increases, as shown in fig. \ref{fig:dynamics} (c) and (d), the expansion of the polariton cloud becomes more prominent: for $\photonweight = 0.66 $ (Fig.~\ref{fig:dynamics}(c)) the polariton cloud reaches a steady state width of $24 \mu$m, while for $\photonweight = 0.82 $ the final width reaches a value of $58  \mu$m (Fig.~\ref{fig:dynamics}(d)).
Interestingly, when examining the data we clearly observe that the timescale over which the expanding distribution reaches the steady state profile is shorter for polaritons which are more photonic, reducing to $\sim 0.7$ ps for $\photonweight = 0.66$ and $\sim 0.3$ ps for $\photonweight = 0.82$. This indicates that not only do polaritons which are more photon-like propagate over longer distances, their expansion also becomes faster.

In addition to the expansion of the polariton cloud, we also observe that the peak of the polariton distribution shifts with time.
In a similar manner to the expansion, the motion of the peak position becomes more pronounced as $\photonweight$ increases (being $\sim 4 \mu$m for $\photonweight = 0.48$ and $\sim 26 \mu$m for $\photonweight = 0.82$ and it appears to stop at the same time as the expansion.
The simultaneous presence of both expansion and peak migration hints at a mixed transport mechanism, governed by both diffusion and ballistic flow.
In fact, the possibility of a mixed transport mechanism is quite plausible, since the two counterparts that make up the polariton exhibit very different transport behavior - Frenkel excitons move in the material in a random, diffusive manner, whereas photons can propagate ballistically over large distances.
Therefore, a fundamental question is - what is the transport nature of the hybrid polaritons?
This question becomes even more intriguing when taking a closer look at the expansion velocities.
While for $\photonweight = 0.82$ the expansion follows the group velocity, which is typical of systems characterized by ballistic transport, this picture changes significantly when the excitonic fraction of the polariton is comparable to the photonic one.
Here, with $\photonweight = 0.48$ (located at point A on the dispersion curve in Fig.~\ref{fig:ss}(c)), the average velocity extracted from the measurements reduces to $\sim$ 7.6 $\mu$m/ps, which is 15 times smaller than the group velocity for those polaritons, calculated to be 109 $\mu$m/ps. 
The reduction in the calculated value of the group velocity is trivial, as the polaritons acquire a larger excitonic fraction and a larger effective mass.
However, the significant discrepancy between the group velocity and the expansion speed extracted from the measurements for those polaritons indicates that the motion becomes non-ballistic, which can be attributed to the underlying disorder in the molecular layer.

\begin{figure}[hp]
\centering
\includegraphics[width=175mm]{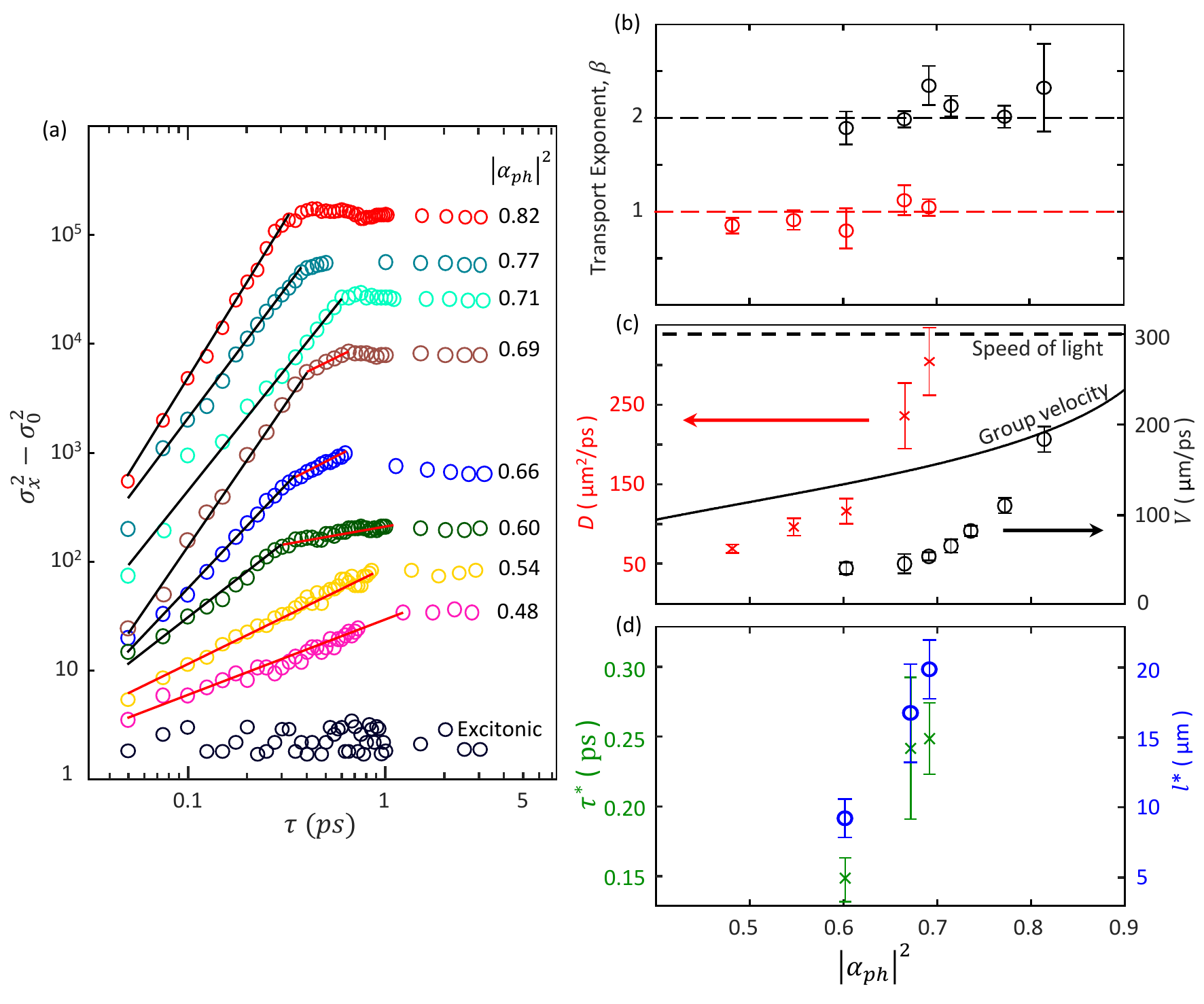}
\caption{
    (a) Time-resolved dynamics of the mean squared displacement (in log-log scale) calculated for the spatial profiles measured for various photonic weights.
    Note that the various plot are offset vertically to aid visibility.
    The solid lines show the results of the linear fits, (corresponding to a power-law relation). The red and black lines indicate a slope close to $\beta=1$ and $\beta=2$, respectively, corresponding to either diffusive or ballistic transport.
    (b) Transport exponent $\beta$ as a function of photonic weight, extracted from the data in (a).
    (c) Transport parameters extracted from the data in (a).
    The values of the diffusion coefficient (red crosses) were obtained from the regions with $\beta \simeq 1$ (marked by red lines in (a)) while the expansion velocity values were obtained from the regions with $\beta \simeq 2$ (black lines in (a)).
    The solid black line shows the theoretical group velocity, calculated from the dispersion of lower BSWP branch and the dashed black line marks the speed of light in vacuum.
    (d) Mean free time (green crosses) and transport mean free path (blue circles), obtained for the transport dynamics for intermediate values of $\photonweight$ at which mixed transport behavior is observed.
}
\label{fig:OVD}
\end{figure}

The changes observed when going between low and high photonic weights points toward a rich behavior which undergoes a transition between different transport regimes.
It is therefore imperative to reveal the various parameters which characterize and govern the transport of polaritons in molecular systems and how those parameters depend on the mixed character of the polaritons.
To answer these questions, we calculate the mean squared displacement (i.e. variance, $\sigma_x^2$) of the polariton distribution probed at various values of $ \photonweight$. We plot $\sigma_x^2$ as function of delay time (in log-log scale) in Fig.~\ref{fig:OVD}(a), which clearly shows one or more distinct linear regions for all values of $ \photonweight$.
The observed linearity over each region matches the general behavior for transport, which is commonly captured by a power-law dependence on time, in the form $\sigma_x^2=\sigma_0^2+D_\beta \tau^\beta$.
Here $D_\beta$ is the generalised diffusion coefficient, $\sigma_0^2$ is the variance at zero delay (which accounts for the point spread function of the of the pump) and $\beta$ is the order of the transport, which is directly linked to the underlying transport mechanism.
A $\beta $ value of unity corresponds to diffusive transport, where the distribution expands as $\sqrt{\tau}$, while $\beta = 2$ represents ballistic transport and expansion at a constant velocity.
A non-integer value of $\beta$ is often associated with anomalous diffusion or transport in fractal systems\cite{Metzler2014}.
To identify the transport nature in each region, we fit the curves shown in Fig.~\ref{fig:OVD}(a) to straight lines (as marked by the black and red lines) and plot the resulting exponents as a function of $\photonweight$ as shown in Fig.~\ref{fig:OVD}(b).
As can be seen, the exponents clearly segregate around two distinct values, $\beta = 1$ or $\beta = 2$.
For $ \photonweight$ values of 0.48 and 0.54, where the the BSWP has almost equal contribution from the excitonic and photonic components, the exponent is close to 1, indicating that the transport is diffusive up to $\tau \sim 1$ ps, at which time the polariton distribution reaches its steady state width.
On the other hand, polaritons with a large photonic fraction ($\photonweight > 0.7$) exhibit a very different behavior, with $\beta$ taking values around 2, indicating that the transport becomes ballistic due to the mixing with the photons and the long-range coherence induced by strong coupling.
Interestingly, at intermediate photonic weights, i.e, with $0.6 \lesssim \photonweight \lesssim 0.7$ the temporal evolution of the polariton cloud shows a crossover between two distinct regions.
For example, for $ \photonweight=0.6$ the expansion is initially ballistic (characterized by a value of $\beta \simeq 2$, indicated by the black line) but at $\tau \simeq 0.3$ ps the exponent changes to a value close to $1$ (shown by the red line), signifying that the transport becomes diffusive.
This diffusive expansion proceed until reaching a steady-state width around $\tau \simeq 1$~ps.
This clearly demonstrates the mixed transport nature of polaritons, which, along with their propagation, experience a transition between ballistic flow and diffusion.
Moreover, a close look at the curves in Fig.~\ref{fig:OVD}(a) reveals that, as the photonic weight of the polariton increases the diffusive region shrinks (e.g. with the transition occurring only at $\tau = \simeq 0.5$~ps for $ \photonweight = 0.69$), until it completely disappears at $ \photonweight > 0.7$.

The data presented in Fig.~\ref{fig:OVD}(a) can be further used to extract the various transport dynamics parameters.
Specifically, in the diffusive regions (corresponding to the red circles with $\beta \simeq 1$ in Fig.~\ref{fig:OVD}(b) or the red lines in Fig.~\ref{fig:OVD}(a)) the diffusion coefficient $D$ can be found by fitting the data to the expression $\sigma_x^2-\sigma_0^2 = 2D \tau$.
The resulting values for the diffusion coefficient as a function of the photonic weight are presented in Fig.~\ref{fig:OVD}(c) (red crosses).
As the the photonic component increases, the diffusion coefficient also increases, from 69~$\mu$m$^2/$ps to 284~$\mu$m$^2/$ps.
This clearly shows that, even though the transport within this parameter range is strongly influenced by the underlying disorder in the system, which makes it non-ballistic, the coherent coupling with the photons and the long-range correlations induced by it play a crucial role in the transport dynamics.
In fact, recalling that the typical diffusion coefficient for excitons in semiconductors (either organic or inorganic) is normally within the range of $10^{-3} - 10^{-7}~\mu$m$^2/$ps ~\cite{Ginsberg2020,Akselrod2014,Zhu2017,Berghuis2021}, our results signify an enhancement of more than six orders of magnitude in the diffusion coefficient.

In a similar manner, we use the data from the ballistic regions (i.e., black circles with $\beta \simeq 2$ in Fig.~\ref{fig:OVD}(b) and black lines in Fig.~\ref{fig:OVD}(a)) and fit it to a linear expansion at a constant velocity, which is shown by the black circles in Fig.~\ref{fig:OVD}(c) as a function of the photonic weight (for $\photonweight \geq 0.6$).
For comparison, we also plot the theoretical group velocity $V_g$ (based on the lower BSWP dispersion), shown by the black solid line, which represents the expected expansion velocity in an ideal, perfectly homogeneous system with fully delocalized wavefunctions.
Once again, the result of this analysis exhibit a clear trend, with the velocity increasing monotonically with $\photonweight$, expressing the interplay between disorder and long-range correlations in the system.
At $\photonweight \sim 0.60$, the lowest photonic weight for which a ballistic region is observed, the expansion velocity is $30 \mu$m/ps.
This velocity is considerably smaller than the group velocity at this photonic weight, which indicates that while the long-range correlations are quite efficient in counteracting the disorder, the effects of the disorder experienced by the excitonic component of the polaritons cannot be ignored.
Once again, this is also expressed by the transition into the diffusive region observed at longer times in Fig.~\ref{fig:OVD}(a).
However, it is interesting to note that even for $\photonweight \gtrsim 0.7$, where the diffusive region disappears and the transport appears to be ballistic throughout the whole expansion, the measured velocity still remains lower than $V_g$.
As the photonic component of the polaritons continue to increase, the effect of the coherent coupling becomes larger, and the measured expansion velocity becomes closer to the theoretical group velocity $V_g$ (for the corresponding value of $\photonweight$).
At the highest photonic weight measured ($\photonweight = 0.82$) the long-range correlations completely overcomes the molecular disorder and the expansion velocity becomes comparable to the group velocity, reaching a value of 176~$\mu$m$/$ps, two thirds the speed of light in vacuum.
The complete mapping of the transport dynamics, as presented in Fig.~\ref{fig:OVD}(a), reveals a very rich, yet simple picture for the motion of polaritons.
Starting with diffusive molecular excitons with a typical mean free path of only several nanometers, the mixing with photons under strong coupling gives rise to extended correlations which increase the transport mean free path.
This, in turn, results in enhanced diffusion, as observed by our measurements.
As the photonic weight is further increased, the mean free path reaches a macroscopic scale (i.e. on the order of 10~$\mu$m), making the ballistic motion between subsequent random scattering events accessible to our time-resolved microscopy.
For a high enough photonic weight, the mean free path becomes comparable to the length scale set by absorption, at which point the system is dominated by dissipation, rather than disorder.
This results in purely ballistic expansion which is now limited by the loss length (which sets the steady-state width of the polariton distribution).
This becomes even more apparent when considering the intermediate cases of $0.6 \leq \photonweight \leq 0.69$ in Fig.~\ref{fig:OVD}(a) and (b) for which we explicitly observe the transition between ballistic and diffusive flow occurring during the propagation.
In fact, since for those specific measurements we obtain, at the same time, both the ballistic velocity and the diffusion coefficient, (see Fig.~\ref{fig:OVD}(c)), we can extract the microscopic quantities governing the transport.
The mean free time ($\tau^\star$) and transport mean free path ($l^\star$) are related to the diffusion coefficient and ballistic velocity by $D=V^2{\tau^\star}/2$ and $l^\star=\frac{\pi}{2} V \tau^\star$ \cite{Datta2018}.
Using the value shown in Fig.~\ref{fig:OVD}(c), we calculate the mean free time and transport mean free path, which are presented in Fig.~\ref{fig:OVD}(d) by the green crosses (for $\tau^\star$) and blue circles (for $l^\star$).
As seen, both parameters increase monotonically with the photonic weight, as expected.
More interestingly, the values obtained for the mean free path are on the order of tens of microns, which is four orders of magnitude larger than the intermolecular distance, expressing the long-range correlation induced by strong coupling.

In conclusion, we presented a comprehensive study of the spatiotemporal dynamics of polaritons in a molecular system strongly coupled to Bloch surface waves.
Taking advantage of the ultra-long propagation of polaritons in such systems, we successfully mapped the transport dynamics and their dependence on the mixing between light and matter.
Our measurements provide access to direct quantification of the microscopic parameters governing the mechanism of cavity-enhanced transport.
We showed that the long-range correlations induced by the coherent coupling with the optical field extend the transport mean free path to macroscopic scales, reaching tens of microns.
As we showed here for the first time, the competition between molecular-scale disorder and long-range correlations leads to mixed transport behavior exhibiting both diffusion and ballistic expansion of polaritons with a clear transition between the two .
This effect initially manifests itself as enhanced diffusion, with the diffusion coefficient reaching values which are six orders of magnitude larger than in bare molecular systems, and, at higher photonic fractions, leads to fully ballistic transport overcoming the molecular disorder.
Finally, we directly showed that the ballistic velocity can reach two thirds the speed of light, even for polaritonic excitation carrying an excitonic fraction of $\sim 20\%$.



\section*{Acknowledgements}
This research was supported by the Israel Science Foundation, grant no. 1435/19 and 1993/13. Tal Schwartz is grateful to Prof. Mordechai Segev for his kind support. 




\bibliography{Transport}

\begin{thebibliography}{10}
\urlstyle{rm}
\expandafter\ifx\csname url\endcsname\relax
  \def\url#1{\texttt{#1}}\fi
\expandafter\ifx\csname urlprefix\endcsname\relax\def\urlprefix{URL }\fi
\expandafter\ifx\csname doiprefix\endcsname\relax\def\doiprefix{DOI: }\fi
\providecommand{\bibinfo}[2]{#2}
\providecommand{\eprint}[2][]{\url{#2}}

\bibitem{Scholes2011}
\bibinfo{author}{Scholes, G.~D.}, \bibinfo{author}{Fleming, G.~R.},
  \bibinfo{author}{Olaya-Castro, A.} \& \bibinfo{author}{{Van Grondelle}, R.}
\newblock \bibinfo{journal}{\bibinfo{title}{{Lessons from nature about solar
  light harvesting}}}.
\newblock {\emph{\JournalTitle{Nature Chemistry}}}
  \textbf{\bibinfo{volume}{3}}, \bibinfo{pages}{763--774}
  (\bibinfo{year}{2011}).

\bibitem{Forrest2020}
\bibinfo{author}{Forrest, S.~R.}
\newblock \emph{\bibinfo{title}{{Organic electronics: Foundations to
  applications}}} (\bibinfo{publisher}{Oxford University Press},
  \bibinfo{year}{2020}).

\bibitem{Tessler2009}
\bibinfo{author}{Tessler, N.}, \bibinfo{author}{Preezant, Y.},
  \bibinfo{author}{Rappaport, N.} \& \bibinfo{author}{Roichman, Y.}
\newblock \bibinfo{journal}{\bibinfo{title}{{Charge transport in disordered
  organic materials and its relevance to thin-film devices: A tutorial
  review}}}.
\newblock {\emph{\JournalTitle{Advanced Materials}}}
  \textbf{\bibinfo{volume}{21}}, \bibinfo{pages}{2741--2761}
  (\bibinfo{year}{2009}).

\bibitem{Ginsberg2020}
\bibinfo{author}{Ginsberg, N.~S.} \& \bibinfo{author}{Tisdale, W.~A.}
\newblock \bibinfo{journal}{\bibinfo{title}{{Spatially Resolved Photogenerated
  Exciton and Charge Transport in Emerging Semiconductors}}}.
\newblock {\emph{\JournalTitle{Annual Review of Physical Chemistry}}}
  \textbf{\bibinfo{volume}{71}}, \bibinfo{pages}{1--30} (\bibinfo{year}{2020}).

\bibitem{Orgiu2015}
\bibinfo{author}{Orgiu, E.} \emph{et~al.}
\newblock \bibinfo{journal}{\bibinfo{title}{{Conductivity in organic
  semiconductors hybridized with the vacuum field}}}.
\newblock {\emph{\JournalTitle{Nature Materials}}}
  \textbf{\bibinfo{volume}{14}}, \bibinfo{pages}{1123--1129}
  (\bibinfo{year}{2015}).

\bibitem{Lerario2017}
\bibinfo{author}{Lerario, G.} \emph{et~al.}
\newblock \bibinfo{journal}{\bibinfo{title}{High-speed flow of interacting
  organic polaritons}}.
\newblock {\emph{\JournalTitle{Light: Science \& Applications}}}
  \textbf{\bibinfo{volume}{6}}, \bibinfo{pages}{e16212--e16212}
  (\bibinfo{year}{2017}).

\bibitem{Rozenman2018}
\bibinfo{author}{Rozenman, G.~G.}, \bibinfo{author}{Akulov, K.},
  \bibinfo{author}{Golombek, A.} \& \bibinfo{author}{Schwartz, T.}
\newblock \bibinfo{journal}{\bibinfo{title}{Long-range transport of organic
  exciton-polaritons revealed by ultrafast microscopy}}.
\newblock {\emph{\JournalTitle{ACS Photonics}}} \textbf{\bibinfo{volume}{5}},
  \bibinfo{pages}{105--110} (\bibinfo{year}{2018}).

\bibitem{Zakharko2018}
\bibinfo{author}{Zakharko, Y.} \emph{et~al.}
\newblock \bibinfo{journal}{\bibinfo{title}{{Radiative Pumping and Propagation
  of Plexcitons in Diffractive Plasmonic Crystals}}}.
\newblock {\emph{\JournalTitle{Nano Letters}}} \textbf{\bibinfo{volume}{18}},
  \bibinfo{pages}{4927--4933} (\bibinfo{year}{2018}).

\bibitem{Nagarajan2020}
\bibinfo{author}{Nagarajan, K.} \emph{et~al.}
\newblock \bibinfo{journal}{\bibinfo{title}{{Conductivity and Photoconductivity
  of a p-Type Organic Semiconductor under Ultrastrong Coupling}}}.
\newblock {\emph{\JournalTitle{ACS Nano}}} \textbf{\bibinfo{volume}{14}},
  \bibinfo{pages}{10219--10225} (\bibinfo{year}{2020}).

\bibitem{Hou2020}
\bibinfo{author}{Hou, S.} \emph{et~al.}
\newblock \bibinfo{journal}{\bibinfo{title}{Ultralong-range energy transport in
  a disordered organic semiconductor at room temperature via coherent
  exciton-polariton propagation}}.
\newblock {\emph{\JournalTitle{Advanced Materials}}}
  \textbf{\bibinfo{volume}{32}}, \bibinfo{pages}{2002127}
  (\bibinfo{year}{2020}).

\bibitem{Bhatt2021}
\bibinfo{author}{Bhatt, P.}, \bibinfo{author}{Kaur, K.} \&
  \bibinfo{author}{George, J.}
\newblock \bibinfo{journal}{\bibinfo{title}{{Enhanced Charge Transport in
  Two-Dimensional Materials through Light-Matter Strong Coupling}}}.
\newblock {\emph{\JournalTitle{ACS Nano}}} \textbf{\bibinfo{volume}{15}},
  \bibinfo{pages}{13616--13622} (\bibinfo{year}{2021}).

\bibitem{Pandya2021}
\bibinfo{author}{Pandya, R.} \emph{et~al.}
\newblock \bibinfo{journal}{\bibinfo{title}{Microcavity-like exciton-polaritons
  can be the primary photoexcitation in bare organic semiconductors}}.
\newblock {\emph{\JournalTitle{Nature Communications}}}
  \textbf{\bibinfo{volume}{12}}, \bibinfo{pages}{6519} (\bibinfo{year}{2021}).

\bibitem{Pandya2022}
\bibinfo{author}{Pandya, R.} \emph{et~al.}
\newblock \bibinfo{journal}{\bibinfo{title}{{Tuning the Coherent Propagation of
  Organic Exciton‐Polaritons through Dark State Delocalization}}}.
\newblock {\emph{\JournalTitle{Advanced Science}}}
  \textbf{\bibinfo{volume}{2}}, \bibinfo{pages}{2105569}
  (\bibinfo{year}{2022}).

\bibitem{Francisco2021}
\bibinfo{author}{Garcia-Vidal, F.~J.}, \bibinfo{author}{Ciuti, C.} \&
  \bibinfo{author}{Ebbesen, T.~W.}
\newblock \bibinfo{journal}{\bibinfo{title}{Manipulating matter by strong
  coupling to vacuum fields}}.
\newblock {\emph{\JournalTitle{Science}}} \textbf{\bibinfo{volume}{373}},
  \bibinfo{pages}{eabd0336} (\bibinfo{year}{2021}).

\bibitem{AberraGuebrou2012}
\bibinfo{author}{{Aberra Guebrou}, S.} \emph{et~al.}
\newblock \bibinfo{journal}{\bibinfo{title}{{Coherent emission from a
  disordered organic semiconductor induced by strong coupling with surface
  plasmons}}}.
\newblock {\emph{\JournalTitle{Physical Review Letters}}}
  \textbf{\bibinfo{volume}{108}}, \bibinfo{pages}{066401}
  (\bibinfo{year}{2012}).

\bibitem{Zhong2017}
\bibinfo{author}{Zhong, X.} \emph{et~al.}
\newblock \bibinfo{journal}{\bibinfo{title}{{Energy Transfer between Spatially
  Separated Entangled Molecules}}}.
\newblock {\emph{\JournalTitle{Angewandte Chemie - International Edition}}}
  \textbf{\bibinfo{volume}{56}}, \bibinfo{pages}{9034--9038}
  (\bibinfo{year}{2017}).

\bibitem{Du2018}
\bibinfo{author}{Du, M.} \emph{et~al.}
\newblock \bibinfo{journal}{\bibinfo{title}{{Theory for polariton-assisted
  remote energy transfer}}}.
\newblock {\emph{\JournalTitle{Chemical Science}}}
  \textbf{\bibinfo{volume}{9}}, \bibinfo{pages}{6659--6669}
  (\bibinfo{year}{2018}).

\bibitem{Georgiou2021}
\bibinfo{author}{Georgiou, K.}, \bibinfo{author}{Jayaprakash, R.},
  \bibinfo{author}{Othonos, A.} \& \bibinfo{author}{Lidzey, D.~G.}
\newblock \bibinfo{journal}{\bibinfo{title}{{Ultralong-Range Polariton-Assisted
  Energy Transfer in Organic Microcavities}}}.
\newblock {\emph{\JournalTitle{Angewandte Chemie International Edition}}}
  \textbf{\bibinfo{volume}{60}}, \bibinfo{pages}{16661--16667}
  (\bibinfo{year}{2021}).

\bibitem{Freixanet2000}
\bibinfo{author}{Freixanet, T.}, \bibinfo{author}{Sermage, B.},
  \bibinfo{author}{Tiberj, A.} \& \bibinfo{author}{Thierry-Mieg, V.}
\newblock \bibinfo{journal}{\bibinfo{title}{Propagation of excitonic polaritons
  in a microcavity}}.
\newblock {\emph{\JournalTitle{physica status solidi (a)}}}
  \textbf{\bibinfo{volume}{178}}, \bibinfo{pages}{133--138}
  (\bibinfo{year}{2000}).

\bibitem{Paravicini-Bagliani2018}
\bibinfo{author}{Paravicini-Bagliani, G.~L.} \emph{et~al.}
\newblock \bibinfo{journal}{\bibinfo{title}{{Magneto-transport controlled by
  Landau polariton states}}}.
\newblock {\emph{\JournalTitle{Nature Physics 2018 15:2}}}
  \textbf{\bibinfo{volume}{15}}, \bibinfo{pages}{186--190}
  (\bibinfo{year}{2018}).

\bibitem{Thomas2019}
\bibinfo{author}{Thomas, A.} \emph{et~al.}
\newblock \bibinfo{title}{{Exploring Superconductivity under Strong Coupling
  with the Vacuum Electromagnetic Field}} (\bibinfo{year}{2019}).
\newblock \eprint{1911.01459}.

\bibitem{Feist2015}
\bibinfo{author}{Feist, J.} \& \bibinfo{author}{Garcia-Vidal, F.~J.}
\newblock \bibinfo{journal}{\bibinfo{title}{{Extraordinary exciton conductance
  induced by strong coupling}}}.
\newblock {\emph{\JournalTitle{Physical Review Letters}}}
  \textbf{\bibinfo{volume}{114}}, \bibinfo{pages}{196402}
  (\bibinfo{year}{2015}).

\bibitem{Schachenmayer2015}
\bibinfo{author}{Schachenmayer, J.}, \bibinfo{author}{Genes, C.},
  \bibinfo{author}{Tignone, E.} \& \bibinfo{author}{Pupillo, G.}
\newblock \bibinfo{journal}{\bibinfo{title}{{Cavity-enhanced transport of
  excitons}}}.
\newblock {\emph{\JournalTitle{Physical Review Letters}}}
  \textbf{\bibinfo{volume}{114}}, \bibinfo{pages}{196403}
  (\bibinfo{year}{2015}).

\bibitem{Gonzalez-Ballestero2016}
\bibinfo{author}{Gonzalez-Ballestero, C.}, \bibinfo{author}{Feist, J.},
  \bibinfo{author}{{Gonzalo Bad{\'{i}}a}, E.}, \bibinfo{author}{Moreno, E.} \&
  \bibinfo{author}{Garcia-Vidal, F.~J.}
\newblock \bibinfo{journal}{\bibinfo{title}{{Uncoupled Dark States Can Inherit
  Polaritonic Properties}}}.
\newblock {\emph{\JournalTitle{Physical Review Letters}}}
  \textbf{\bibinfo{volume}{117}}, \bibinfo{pages}{156402}
  (\bibinfo{year}{2016}).

\bibitem{Hagenmuller2017}
\bibinfo{author}{Hagenm{\"{u}}ller, D.}, \bibinfo{author}{Schachenmayer, J.},
  \bibinfo{author}{Sch{\"{u}}tz, S.}, \bibinfo{author}{Genes, C.} \&
  \bibinfo{author}{Pupillo, G.}
\newblock \bibinfo{journal}{\bibinfo{title}{{Cavity-Enhanced Transport of
  Charge}}}.
\newblock {\emph{\JournalTitle{Physical Review Letters}}}
  \textbf{\bibinfo{volume}{119}}, \bibinfo{pages}{223601}
  (\bibinfo{year}{2017}).

\bibitem{Sentef2018}
\bibinfo{author}{Sentef, M.~A.}, \bibinfo{author}{Ruggenthaler, M.} \&
  \bibinfo{author}{Rubio, A.}
\newblock \bibinfo{journal}{\bibinfo{title}{{Cavity quantum-electrodynamical
  polaritonically enhanced electron-phonon coupling and its influence on
  superconductivity}}}.
\newblock {\emph{\JournalTitle{Science Advances}}} \textbf{\bibinfo{volume}{4}}
  (\bibinfo{year}{2018}).

\bibitem{Botzung2020}
\bibinfo{author}{Botzung, T.} \emph{et~al.}
\newblock \bibinfo{journal}{\bibinfo{title}{{Dark state semilocalization of
  quantum emitters in a cavity}}}.
\newblock {\emph{\JournalTitle{Physical Review B}}}
  \textbf{\bibinfo{volume}{102}}, \bibinfo{pages}{144202}
  (\bibinfo{year}{2020}).

\bibitem{Chavez2021}
\bibinfo{author}{Ch{\'{a}}vez, N.~C.}, \bibinfo{author}{Mattiotti, F.},
  \bibinfo{author}{M{\'{e}}ndez-Berm{\'{u}}dez, J.~A.},
  \bibinfo{author}{Borgonovi, F.} \& \bibinfo{author}{Celardo, G.~L.}
\newblock \bibinfo{journal}{\bibinfo{title}{{Disorder-Enhanced and
  Disorder-Independent Transport with Long-Range Hopping: Application to
  Molecular Chains in Optical Cavities}}}.
\newblock {\emph{\JournalTitle{Physical Review Letters}}}
  \textbf{\bibinfo{volume}{126}} (\bibinfo{year}{2021}).

\bibitem{Engelhardt2022}
\bibinfo{author}{Engelhardt, G.} \& \bibinfo{author}{Cao, J.}
\newblock \bibinfo{journal}{\bibinfo{title}{{Unusual dynamical properties of
  disordered polaritons in microcavities}}}.
\newblock {\emph{\JournalTitle{Physical Review B}}}
  \textbf{\bibinfo{volume}{105}}, \bibinfo{pages}{64205}
  (\bibinfo{year}{2022}).

\bibitem{Virgili2012}
\bibinfo{author}{Virgili, T.} \emph{et~al.}
\newblock \bibinfo{journal}{\bibinfo{title}{{Confocal ultrafast pump–probe
  spectroscopy: a new technique to explore nanoscale composites}}}.
\newblock {\emph{\JournalTitle{Nanoscale}}} \textbf{\bibinfo{volume}{4}},
  \bibinfo{pages}{2219} (\bibinfo{year}{2012}).

\bibitem{Akselrod2014}
\bibinfo{author}{Akselrod, G.~M.} \emph{et~al.}
\newblock \bibinfo{journal}{\bibinfo{title}{Visualization of exciton transport
  in ordered and disordered molecular solids}}.
\newblock {\emph{\JournalTitle{Nature Communications}}}
  \textbf{\bibinfo{volume}{5}}, \bibinfo{pages}{3646} (\bibinfo{year}{2014}).

\bibitem{Zhu2017}
\bibinfo{author}{Zhu, T.}, \bibinfo{author}{Wan, Y.} \& \bibinfo{author}{Huang,
  L.}
\newblock \bibinfo{journal}{\bibinfo{title}{{Direct Imaging of Frenkel Exciton
  Transport by Ultrafast Microscopy}}}.
\newblock {\emph{\JournalTitle{Accounts of Chemical Research}}}
  \textbf{\bibinfo{volume}{50}}, \bibinfo{pages}{1725--1733}
  (\bibinfo{year}{2017}).

\bibitem{Wan2015}
\bibinfo{author}{Wan, Y.} \emph{et~al.}
\newblock \bibinfo{journal}{\bibinfo{title}{{Cooperative singlet and triplet
  exciton transport in tetracene crystals visualized by ultrafast
  microscopy}}}.
\newblock {\emph{\JournalTitle{Nature Chemistry}}}
  \textbf{\bibinfo{volume}{7}}, \bibinfo{pages}{785--792}
  (\bibinfo{year}{2015}).

\bibitem{Delor2020}
\bibinfo{author}{Delor, M.}, \bibinfo{author}{Weaver, H.~L.},
  \bibinfo{author}{Yu, Q.~Q.} \& \bibinfo{author}{Ginsberg, N.~S.}
\newblock \bibinfo{journal}{\bibinfo{title}{{Imaging material functionality
  through three-dimensional nanoscale tracking of energy flow}}}.
\newblock {\emph{\JournalTitle{Nature Materials}}}
  \textbf{\bibinfo{volume}{19}}, \bibinfo{pages}{56--62}
  (\bibinfo{year}{2020}).

\bibitem{Berghuis2021}
\bibinfo{author}{Berghuis, A.~M.} \emph{et~al.}
\newblock \bibinfo{journal}{\bibinfo{title}{{Effective Negative Diffusion of
  Singlet Excitons in Organic Semiconductors}}}.
\newblock {\emph{\JournalTitle{Journal of Physical Chemistry Letters}}}
  \textbf{\bibinfo{volume}{12}}, \bibinfo{pages}{1360--1366}
  (\bibinfo{year}{2021}).

\bibitem{Yeh1978}
\bibinfo{author}{Yeh, P.}, \bibinfo{author}{Yariv, A.} \& \bibinfo{author}{Cho,
  A.~Y.}
\newblock \bibinfo{journal}{\bibinfo{title}{{Optical surface waves in periodic
  layered media}}}.
\newblock {\emph{\JournalTitle{Applied Physics Letters}}}
  \textbf{\bibinfo{volume}{32}}, \bibinfo{pages}{104--105}
  (\bibinfo{year}{1978}).

\bibitem{Lerario2014}
\bibinfo{author}{Lerario, G.} \emph{et~al.}
\newblock \bibinfo{journal}{\bibinfo{title}{{Room temperature Bloch surface
  wave polaritons}}}.
\newblock {\emph{\JournalTitle{Optics Letters}}} \textbf{\bibinfo{volume}{39}},
  \bibinfo{pages}{2068} (\bibinfo{year}{2014}).

\bibitem{Ohad2018}
\bibinfo{author}{Ohad, A.} \emph{et~al.}
\newblock \bibinfo{journal}{\bibinfo{title}{{Spatially resolved measurement of
  plasmon dispersion using Fourier-plane spectral imaging}}}.
\newblock {\emph{\JournalTitle{Photonics Research}}}
  \textbf{\bibinfo{volume}{6}}, \bibinfo{pages}{653} (\bibinfo{year}{2018}).

\bibitem{Schwartz2013}
\bibinfo{author}{Schwartz, T.} \emph{et~al.}
\newblock \bibinfo{journal}{\bibinfo{title}{Polariton dynamics under strong
  light-molecule coupling}}.
\newblock {\emph{\JournalTitle{ChemPhysChem}}} \textbf{\bibinfo{volume}{14}},
  \bibinfo{pages}{125--131} (\bibinfo{year}{2013}).

\bibitem{Fischer2016}
\bibinfo{author}{Fischer, M.~C.}, \bibinfo{author}{Wilson, J.~W.},
  \bibinfo{author}{Robles, F.~E.} \& \bibinfo{author}{Warren, W.~S.}
\newblock \bibinfo{journal}{\bibinfo{title}{{Invited Review Article: Pump-probe
  microscopy}}}.
\newblock {\emph{\JournalTitle{Review of Scientific Instruments}}}
  \textbf{\bibinfo{volume}{87}}, \bibinfo{pages}{031101}
  (\bibinfo{year}{2016}).

\bibitem{Mewes2020}
\bibinfo{author}{Mewes, L.}, \bibinfo{author}{Wang, M.},
  \bibinfo{author}{Ingle, R.~A.}, \bibinfo{author}{B{\"{o}}rjesson, K.} \&
  \bibinfo{author}{Chergui, M.}
\newblock \bibinfo{journal}{\bibinfo{title}{{Energy relaxation pathways between
  light-matter states revealed by coherent two-dimensional spectroscopy}}}.
\newblock {\emph{\JournalTitle{Communications Physics}}}
  \textbf{\bibinfo{volume}{3}}, \bibinfo{pages}{1--10} (\bibinfo{year}{2020}).

\bibitem{Metzler2014}
\bibinfo{author}{Metzler, R.}, \bibinfo{author}{Jeon, J.~H.},
  \bibinfo{author}{Cherstvy, A.~G.} \& \bibinfo{author}{Barkai, E.}
\newblock \bibinfo{journal}{\bibinfo{title}{{Anomalous diffusion models and
  their properties: Non-stationarity, non-ergodicity, and ageing at the
  centenary of single particle tracking}}}.
\newblock {\emph{\JournalTitle{Physical Chemistry Chemical Physics}}}
  \textbf{\bibinfo{volume}{16}}, \bibinfo{pages}{24128--24164}
  (\bibinfo{year}{2014}).

\bibitem{Datta2018}
\bibinfo{author}{Datta, S.}
\newblock \bibinfo{title}{Part b: Quantum transport}.
\newblock In \emph{\bibinfo{booktitle}{Lessons from Nanoelectronics:A New
  Perspective on Transport}}, vol.~\bibinfo{volume}{05},
  \bibinfo{pages}{43--49} (\bibinfo{publisher}{World Scientific},
  \bibinfo{year}{2018}).

\end{thebibliography}

\end{document}